\documentstyle[12pt]{article}
\renewcommand\baselinestretch{1.15}
\advance\textheight by \topskip
\begin{document}

\def\lsim{\mathrel{\lower2.5pt\vbox{\lineskip=0pt\baselineskip=0pt
					 \hbox{$<$}\hbox{$\sim$}}}}
\def\gsim{\mathrel{\lower2.5pt\vbox{\lineskip=0pt\baselineskip=0pt
					 \hbox{$>$}\hbox{$\sim$}}}}
\def\gs{SU(2)_{\rm L} \times U(1)_{\rm Y}}
\def\wh{\hat{W}}
\def\bh{\hat{B}}
\def\wtu{\hat{W}^{\mu \nu}}
\def\wtd{\hat{W}_{\mu \nu}}
\def\btu{\hat{B}^{\mu \nu}}
\def\btd{\hat{B}_{\mu \nu}}
\def\cl{{\cal L}}
\def\nll{\cl_{\rm NL}}
\def\ecl{\cl_{\rm EChL}}
\def\fpnl{\cl_{\rm FP}^{\rm NL}}
\def\fpl{\cl_{\rm FP}}
\def\msb{{\overline{\rm MS}}}
\def\mh{M_H}

\begin{titlepage}

\title{\LARGE \bf  The Electroweak Chiral Lagrangian for the \\
Standard Model with a heavy Higgs}

\author{ {\bf Maria J. Herrero}\thanks{e--mail:herrero@vm1.sdi.uam.es}\
\ \ {\bf and \  Ester Ruiz Morales}\thanks{e--mail:
meruiz@vm1.sdi.uam.es, meruiz@slacvm.bitnet} \\[1mm]
Departamento de F\'{\i}sica
Te\'orica\\ Universidad Aut\'onoma de Madrid\\ Cantoblanco
\ \ 28049 -- Madrid\ \ Spain}
\date{}
\maketitle

\begin{abstract}
\def\baselinestretch{1.3}
\noindent
{\large The most general chiral Lagrangian for electroweak interactions
with the complete set of $SU(2)_L\times U(1)_Y$ invariant operators up
to dimension four is considered. The two-point and three-point functions
with external gauge fields are derived from this effective chiral
Lagrangian to one-loop order in a generic $R_\xi$-gauge. The same set of
Green's functions are paralelly studied in the renormalizable standard
model to one-loop order, in a $R_\xi$-gauge and in the large Higgs mass
limit. An appropriate set of matching conditions connecting the Green's
functions of the two theories allows us to derive, systematically,
the values of the chiral Lagrangian coefficients corresponding to the
large Higgs mass limit of the standard model. These  chiral
parameters represent the non-decoupling effects of a heavy Higgs
particle and incorporate both the leading logarithmic dependence on
$\mh$ and the next to leading constant contributions.
Some phenomenological implications are also discussed.}
\end{abstract}

\vskip-20cm
\rightline{{\bf \large FTUAM 93/24}}
\rightline{{\bf \large July 1993}}

\end{titlepage}

\newpage
\def\baselinestretch{1.25}

\section{Introduction}

Chiral Lagrangians have been extensively used to describe the
phenomenon of spontaneous symmetry breaking both in strong and
electroweak interactions.  The basis of this approach was formulated
by Weinberg \cite{W} to characterize the most general S-matrix
elements for soft pion interactions; and later on it was greatly
developed by Gasser and Leutwyler \cite{GL} in a well defined
framework called chiral perturbation theory  describing low energy
aspects of strong interactions \cite{D}.  The use of chiral Lagrangians
as low energy effective theories for electroweak interactions has
received much attention in the past few years
\cite{DHD,DHT,HDG,EH}.	The electroweak chiral Lagrangian provides
the most general parametrization of the Higgs phenomenon for the
spontaneous breaking of the $\gs$ symmetry in terms of the minimum
number of degrees of freedom, namely, the three would-be Goldstone
bosons (GB). These bosons are incorporated into non-linear
representations of the $\gs$ symmetry group such that the electroweak
chiral Lagrangian built up with these modes and the gauge fields is
manifestly $\gs$ invariant.  The price to be paid in this Higgs-less
parametrization is that the resulting low energy theory is
non-renormalizable, and a tower of new counterterms of increasing
dimension has to be added at each loop order to render the theory
finite.

{}From the point of view	of effective field theories, chiral
Lagrangians can be regarded as the low energy limit of an underlying
fundamental theory, where some heavy fields have been integrated out
inducing additional higher dimension operators.  These effective
operators  can, in principle, be determined if the underlying
fundamental interactions are known.  In a perturbative approach, it
is done by explicit calculation of the relevant loop diagrams and by
matching the predictions of the full underlying theory (in which
heavy particles are present) and those of the low energy effective
theory (with only light degrees of freedom) at some reference scale
\cite{G}.  This combined picture of integrating out the heavy fields
and matching the predictions of the two theories
has been recently applied to some particular situations
in electroweak interactions. One example is provided in the standard
model when the heavy top quark is integrated out at one loop level
\cite{FMM}.  Other situations when going beyond the standard model
have also been considered \cite{F,Sint}. Technicolor models
\cite{TEC} where one or more
pairs of techniquark doublets are integrated out are typical examples
\cite{HDG,Lynn,AW}.

In this paper, we have considered the possibility where the standard
model itself is the underlying theory and the heavy field to be
integrated out is the Higgs particle.  Our aim is to determine the
electroweak chiral Lagrangian (EChL), being $\gs$ symmetric, that
parametrizes electroweak interactions to one loop order in the case
of a heavy Higgs particle. We will deduce the values of the EChL
parameters by integrating out the Higgs field to one loop
level and by matching the standard model predictions in the large
$\mh$ limit with the predictions from the chiral Lagrangian to one
loop order.  By
large $\mh$ limit we mean the situation where the mass of the Higgs
is much larger than the available external momenta (p$^2 \ll \mh^2$)
and other particle masses (m$^2 \ll \mh^2$, m= M$_{\rm W}$, M$_{\rm Z}$),
 but not so high that perturbation theory is unreliable ($\mh^2 \lsim 1
$TeV$^2$).

The large $\mh$ limit of the standard model has been studied before
by several authors in the context of the gauged non-linear sigma
model. In the pioneering works of Appelquist and Bernard \cite{AB}
and Longhitano \cite{L}, all the new divergences appearing  when
calculating with the non-renormalizable non-linear sigma model to one
loop order were sistematically found. By using the fact that the
mass of the Higgs particle acts as a regulator for the standard
model, they could identify these new divergences with the logarithmic
dependence on $\mh$ of the observables in the standard model
(see also \cite{VG}).
However, with this approach, one cannot determine the finite
(non-logarithmic) Higgs contributions to the EChL parameters that are
of the same order in the perturbative expansion.  In contrast, the
effective chiral Lagrangian approach that we propose to use here
reproduces correctly the two kind of Higgs contributions.  As we will
show in this work by explicit computation of the two and three
point functions for gauge bosons to one loop, there are indeed finite
contributions to the EChL parameters that, together with the
logarithmic contributions, account for the complete non-decoupling
effects of the Higgs particle in the large $\mh$ limit.

In section 2, we present the complete $\gs$ invariant electroweak chiral
Lagrangian and fix the prescription to calculate renormalized
Green's functions with this Lagrangian to one loop order. The set of
matching conditions relating the Green's functions
obtained from the effective theory to one loop with the Green's functions
computed in the Standard Model to one loop and in the large $\mh$ limit
will be established in section 3. The values for the EChL
parameters corresponding to a heavy Higgs particle of the Standard Model
will also be derived in that section. Section 4 is devoted to some
discussions on	the most relevant phenomenological consequences of our
results. The conclusions are summarized in section 5.

\section{The electroweak chiral Lagrangian}
We start by writing down the electroweak chiral Lagrangian that is
$\gs$ invariant and contains the whole set of CP-invariant operators
up to dimension four. We will use the notation of ref. \cite{F} that
is closely related to Longhitano's notation, and will restrict our
study to the bosonic sector of electroweak theory.  The basic
building blocks that are used in the construction of the Lagrangian
are the following:
\begin{eqnarray}
T & \equiv & U \tau^3 U^\dagger \nonumber\\[2mm] V_\mu & \equiv &
(D_\mu U) U^\dagger \nonumber\\[2mm] D_\mu U & \equiv & \partial_\mu
U - g \wh_\mu U + g' U \bh_\mu \nonumber\\[2mm]
\wtd & \equiv &  \partial_\mu \wh_\nu - \partial_\nu \wh_\mu -
g [\wh_\mu, \wh_\nu]\nonumber\\[2mm]
\btd & \equiv &  \partial_\mu \bh_\nu - \partial_\nu \bh_\mu,
\end{eqnarray}
where the light bosonic fields have been parametrized as
\begin{eqnarray}
U & \equiv & \displaystyle{\exp\left( {i \;
\frac{\vec{\tau}\cdot\vec{\pi}}{v}}\right)},
\;\;\; v  = 246 \;{\rm GeV}, \;\;\; \vec{\pi} = (\pi^1,\pi^2,\pi^3),
\nonumber\\ \wh_\mu & \equiv & \frac{ -i}{2}\; \vec{W}_\mu \cdot
\vec{\tau},  \nonumber\\ \bh_\mu & \equiv & \frac{ -i}{2} \; B_\mu \;
\tau^3. \end{eqnarray}
Their transformation properties under $\gs$ are
\begin{eqnarray}
\wh'_\mu & = & g_{\rm L} \wh_\mu g_{\rm L}^\dagger - \frac{1}{g} g_{\rm L}
\partial_\mu
g_{\rm L}^\dagger\nonumber\\[2mm]
\bh'_\mu & = &  \bh_\mu  - \frac{1}{g'} g_{\rm R}  \partial_\mu g_{\rm
R}^\dagger
\end{eqnarray}
where
\begin{equation}
g_{\rm L} = e^{\displaystyle{ i \vec{\alpha}\cdot\vec{\tau} / 2}} \;\;\in
\; SU(2)_{\rm L} , \;\;\;\; g_{\rm R} = e^{\displaystyle{ i \beta \tau^3 / 2}}
\;\;\in \; U(1)_Y .
\end{equation}
The would-be Goldstone boson fields $\vec{\pi}$ transform
non-linearly whereas the $U$ field transforms linearly
\begin{eqnarray}
U' & = & g_{\rm L} U g_{\rm R}^\dagger ,\nonumber\\[2mm]
\vec{\pi'} \cdot \vec{\tau} & = & \vec{\pi} \cdot \vec{\tau} +
 v  \vec{\alpha} \cdot \frac{\vec{\tau}}{2} - v \beta
\frac{\tau^3}{2} - (\vec{\alpha}\times\vec{\pi})\cdot
\frac{\vec{\tau}}{2} + \frac{\beta}{2}(\pi_2 \tau_1-\pi_1 \tau_2)+
\nonumber\\[2mm]
 & &
+\frac{1}{6v}[(\vec{\alpha}\cdot\vec{\pi})
(\vec{\pi}\cdot\vec{\tau})-(\vec{\alpha}\cdot\vec{\tau})
(\vec{\pi}\cdot\vec{\pi})]-\frac{\beta}{6v}
[\pi_3(\vec{\pi}\cdot\vec{\tau})-\tau_3(\vec{\pi}\cdot\vec{\pi})]
 + O(\pi^3) .
\end{eqnarray}
The physical gauge fields are defined as
\begin{eqnarray}
W^\pm_\mu & = & \frac{W^1_\mu \mp i W^2_\mu}{\sqrt 2},
\nonumber\\[2mm] Z_\mu & = & c_{\rm w} W^3_\mu - s_{\rm w} B_\mu \nonumber,
\\[2mm] A_\mu & = & s_{\rm w} W^3_\mu + c_{\rm w} B_\mu,
\end{eqnarray}
where $c_{\rm w} = \cos \theta_{\rm w}, \; s_{\rm w} = \sin \theta_{\rm w}$ and
the weak
angle is defined by $\tan \theta_{\rm w} = g' / g$.

At  tree level and to lowest order in the derivative expansion,
the effective low energy theory for the standard model with a heavy
Higgs is the well known Lagrangian of the gauged non-linear sigma
model, given by \cite{L}:
\begin{eqnarray}
 \nll & = & \frac{v^2}{4}\; Tr\left[ D_\mu U^\dagger D^\mu U
 \right] + \cl_{\rm G},
\label{NLL} \\[2mm]
\cl_{\rm G} & = & \frac{1}{2}\; Tr\left[ \wtd \wtu + \btd \btu \right] +
\cl_{\rm GF} + \fpnl.
\nonumber
\end{eqnarray}
We have chosen to work in an arbitrary R$_\xi$-covariant gauge with
the following gauge fixing term
\begin{equation}
\cl_{\rm GF} =-\frac{1}{2\xi_B}f_0^2-
\frac{1}{2\xi_W} \left( \sum_{i=1}^3 f_i^2 \right) ,
\end{equation}
where
\begin{eqnarray}
f_0 & = &   \partial_\mu B^\mu +
\frac{g' v \xi_B}{2} \ \pi_3 ,\nonumber \\[2mm]
f_i & = &   \partial_\mu W_i^\mu -
\frac{g v \xi_W}{2} \  \pi_i , \;\;\; i=1,2,3.
\end{eqnarray}
and the corresponding Faddeev--Popov ghost term is given by
\begin{equation}
 \fpnl =\sum_{a,b=0}^{3} c_a^\dagger (x) \frac{\delta f_a}{\delta
 \theta_b} c_b(x) ,
\end{equation}
where $\theta_0 = \beta, \, \theta_i = \alpha_i, \, i=1,2,3$.

It is important to menction that $\fpnl$ does not
coincide with the usual Faddeev--Popov Lagrangian of the standard
model.	Due to the non-linearity of the
would-be Goldstone boson modes under infinitesimal $\gs$ transformations,
new interactions with increasing number of GB will appear in $\fpnl$.
An explicit expression for $\fpnl$ is given in Appendix A.
The relevant result is
that the couplings of the type two ghosts--one gauge boson and two
ghosts--one GB remain the same as in the linear model; but in
addition there are new couplings of the type two ghosts--two or more
GB that replace the standard coupling two ghosts--one Higgs. Since all
these new couplings are proportional to the $\xi$-parameter, it turns
out that when the Landau gauge is chosen ($\xi=0$), the
Faddeev--Popov terms in the EChL, $\fpnl$,  and the standard
model, $\fpl$, coincide.

The convenience of the Landau gauge choice in the context of the
gauged non-linear sigma model was emphasized by Appelquist and
Bernard in \cite{AB}. Since in this gauge there are no direct
copulings of the GB to the ghosts, the non-renormalizability of
the GB self-interactions in $\nll$ does not infect graphs
 with external ghosts. For this reason, the counterterms
needed to cancel the divergences generated with $\nll$ at one loop
are gauge invariant functions of the GB and gauge fields only.
In other R$_\xi$--gauges, there will be also counterterms
that are functions of the ghost fields	and their structure will
have to be determined by using the more general Becchi-Rouet-Stora
invariance.  This fact, of course, does not preclude the use of
R$_\xi$--gauges, but rather establishes the Landau gauge as the
easiest one in the context of chiral electroweak theories.
Alternatively, if one calculates only Green's fuctions with
external gauge particles $\gamma, W^\pm$ and $Z$ as in the present work,
all the new required counterterms are manifestly gauge invariant local
functions of the GB and gauge fields. This fact can be seen by
computing the Green's functions in a generic R$_\xi$--gauge and by
checking explicitely the independence of the new counterterms on the
$\xi$ parameter.

The non-linear Lagrangian of eq.(\ref{NLL}), when treated at tree level,
reproduces correctly  the low energy properties
of the standard model at tree level and in the heavy Higgs mass
limit. In particular, it leads to the proper low energy
theorems for the scattering amplitudes of longitudinal vector
bosons \cite{LE}. However, in order to incorporate in the low energy
theory
the effects of heavy particles beyond the tree level, new effective
operators have to be considered.  The complete electroweak  chiral
Lagrangian \cite{L} containing the whole set of $\gs$ and CP-
invariant operators up to dimension four is the following\footnote{
There is an extra term $\cl_{14}$ proportional to $\epsilon^{\mu
\nu\alpha\beta}$ that is CP conserving but C and P violating
\cite{F,AW}.
It is not relevant in case of absence of fermion contributions and
will not be considered here.}
\begin{equation}
\ecl = \cl_{\rm   NL} + \sum_{i=0}^{13} \cl_{i} ,\label{ECL}
\end{equation}
where $\nll$ is given in eq.(\ref{NLL}) and the new
operators are\footnote{ The relation with Longhitano's notation is the
following: $ a_0=g^2\beta_1\;;\;a_1=\frac{g}{g'}\alpha_1\;;\;
a_2=\frac{g}{g'}\alpha_2\;;\;a_3=-\alpha_3\;;\;a_i=\alpha_i\;,
i=4,5,6,7\;;\;a_8=-\alpha_8\;;\;a_9=-\alpha_9\;;\;
a_{10}=\alpha_{10}/2\;;\;a_{11}=\alpha_{11}\;;\;
a_{12}=\alpha_{12}/2\;;\;a_{13}=\alpha_{13}$.}:
\begin{eqnarray}
\cl_{0} & = & a_0 \frac{v^2}{4} \left[ Tr\left( T V_\mu \right) \right]^2
\nonumber\\[2mm]
\cl_{1} & = & a_1 \frac{i g g'}{2} B_{\mu\nu}  Tr\left( T \wtu \right)
\nonumber\\[2mm]
\cl_{2} & = & a_2 \frac{i g'}{2} B_{\mu\nu} Tr\left( T [V^\mu,V^\nu ]
\right)
\nonumber\\[2mm]
\cl_{3} & = & a_3  g Tr\left( \wtd [V^\mu,V^\nu ]\right)
\nonumber\\[2mm]
\cl_{4} & = & a_4  \left[ Tr\left( V_\mu V_\nu \right) \right]^2
\nonumber\\[2mm]
\cl_{5} & = & a_5  \left[ Tr\left( V_\mu V^\mu \right) \right]^2
\nonumber\\[2mm]
\cl_{6} & = & a_6 Tr\left( V_\mu V_\nu \right) Tr\left( T V^\mu
\right) Tr\left( T V^\nu \right)\nonumber\\[2mm]
\cl_{7} & = & a_7 Tr\left( V_\mu V^\mu \right) \left[ Tr\left( T V^\nu \right)
\right]^2\nonumber\\[2mm]
\cl_{8} & = & a_8  \frac{g^2}{4} \left[ Tr\left( T \wtd \right) \right]^2
\nonumber\\[2mm]
\cl_{9} & = & a_9  \frac{g}{2} Tr\left( T \wtd \right) Tr\left( T [V^\mu,V^\nu
]
\right) \nonumber\\[2mm]
\cl_{10} & = & a_{10} \left[ Tr\left( T V_\mu \right) Tr\left( T V_\nu ]
\right)
\right]^2 \nonumber\\[2mm]
\cl_{11} & = & a_{11} Tr\left( ( D_\mu V^\mu )^2 \right) \nonumber\\[2mm]
\cl_{12} & = & a_{12} Tr\left( T D_\mu D_\nu V^\nu \right) Tr \left( T V^\mu
\right)\nonumber\\[2mm]
\cl_{13} & = & a_{13} \frac{1}{2} \left[ Tr \left( T D_\mu V_\nu \right)
\right]^2 \label{Li}
\end{eqnarray}
Making use of the equations of motion, the above list can be reduced
to eleven independent operators. In particular, one could eliminate
the 11, 12 and 13 terms by redefining the rest of the terms \cite{F}.
However, since we will not restrict ourselves to calculate on-shell
matrix elements, we keep the complete basis given in eq.(\ref{Li}).
Among all these 14 terms, only the first six
are truly needed as counterterms to absorb the new divergent
structures that appear when calculating to one loop with the
non-renormalizable Lagrangian $\nll$ of eq.(\ref{NLL}).
However, we  retain the complete set to
parametrize all the effects of a heavy Higgs to one loop order
including both finite and divergent contributions.

We will now give the prescription to compute finite renormalized 1PI
Green's functions $\Gamma_{\rm R}^{\rm EChL}$, to one loop order with the
effective chiral electroweak Lagrangian of eq.(\ref{ECL}). We choose
to regulate the divergences by means of the dimensional
regularization method that preserves the Ward identities of the
effective theory\footnote{ A recent discussion on regularization methods
in Chiral Perturbation Theory has been done in \cite{EM}}.

Firstly,
there will be contributions to
$\Gamma_{\rm R}^{\rm EChL}$ from $\nll$ and the $\cl_i$ 's when
used at tree level. These terms will also act as source of
counterterms when we rescale the fields and parameters in $\ecl$
according to \cite{L}: \begin{eqnarray}
B_{\mu}^b & = & \widehat{Z}_B^{1/2} B_\mu \nonumber\\[2mm]
\vec{W}_\mu^b & = & \widehat{Z}_W^{1/2} \vec{W}_\mu
\nonumber\\[2mm] \vec{\pi}^b & =& \widehat{Z}_\pi^{1/2}
\vec{\pi} \nonumber\\[2mm] g^b & =& \widehat{Z}_W^{-1/2} ( g -
\widehat{\delta g} )\nonumber\\[2mm] g'^b & = &
\widehat{Z}_B^{-1/2} ( g' - \widehat{\delta g'} )\nonumber\\[2mm]
v^b & = & \widehat{Z}_\pi^{1/2} ( v -\widehat{ \delta v})\nonumber\\[2mm]
\xi_B^b & = & \xi_B ( 1 + \widehat{\delta \xi}_B)\nonumber\\[2mm]
\xi_W^b & = & \xi_W ( 1 + \widehat{\delta \xi}_W)
\end{eqnarray}
where $ \widehat{Z}_i \equiv 1 + \widehat{\delta Z_i}$ and the
superscript b denotes bare quantities. The fields and parameters
appearing in the right hand side of eq.(13) are renormalized
quantities.  We have used the hatted notation to distinguish the
counterterms of the effective theory from the corresponding quantities in
the  standard model, to be presented in the next section.
Similarly, we rescale the $a_i$'s according to:
\begin{equation} a_i^b = a_i + \delta a_i
\end{equation}
where the $a_i$'s in the right hand side are renormalized parameters.

Secondly, there will be contributions to $\Gamma_{\rm R}^{\rm EChL}$ from
loops generated by $\nll$ and the $\cl_i$'s. The
contributions from the $\cl_i$'s at one loop level are subleading
with respect to the corresponding ones from $\nll$ because they
are either of higher order in powers of the gauge couplings or of
higher order in powers of the external momenta, so that they will be
neglected from now on.

Finally the renormalized 1PI Green's functions with external gauge
particles can be formally defined as
\begin{equation}
\Gamma_{\rm R}^{\rm EChL}(\mu) = \Gamma_0^{\rm EChL} + \Gamma_{\rm C}^{\rm
EChL} + \Gamma_{\rm L}^{\rm EChL}
\end{equation}
where $\Gamma_0^{\rm EChL}$ is the contribution from $\ecl$
at the tree level, $\Gamma_{\rm C}^{\rm EChL}$ is the contribution
from the counterterms and $\Gamma_{\rm L}^{\rm EChL}$ is the contribution
from the loops generated by $\nll$. The energy scale $\mu$ is the usual
 scale of dimensional regularization. The final result for
$\Gamma_{\rm R}^{\rm EChL}$ must be
expressed in terms of the renormalized parameters.

With this prescription at hand one is able to give a finite result
for $\Gamma_{\rm R}^{\rm EChL}$.  The new divergences generated by
$\nll$ at one loop are absorbed into the
redefinitions of the $a_i$'s such that the final result for
$\Gamma_{\rm R}^{\rm EChL}$ is given in terms of the renormalized
parameters $ a_i(\mu) = a_i^b - \delta a_i$  that are, in general,
$\mu$-scale and renormalization prescription dependent.  These
renormalized $a_i(\mu)$ parameters incorporate the
effects of the heavy particles at one loop, and can be explicitely
computed by the matching procedure when the underlying fundamental
theory is known.

\section{Effective Lagrangian parameters for a heavy Higgs}

As we have already said, the fundamental theory that we want to
represent by the electroweak chiral Lagrangian is the standard model
with a heavy Higgs particle.  We start with the standard model
Lagrangian
\begin{equation}
\cl_{\rm SM} = (D_\mu \Phi)^\dagger (D^\mu \Phi) + \mu^2 \Phi^\dagger \Phi -
\lambda (\Phi^\dagger \Phi)^2 + \frac{1}{2} Tr \left( \wtd \wtu + \btd \btu
\right)
+ \cl_{\rm GF} + \cl_{\rm FP} ,
\end{equation}
where
\begin{equation}
\Phi  = \frac{1}{\sqrt 2}\left( \begin{array}{c}\phi_1 - i \phi_2 \\
	 \sigma + i \chi \end{array}\right), \;\;\;\;\;
(\pi_1,\pi_2,\pi_3)  \equiv  (-\phi_2,\phi_1,-\chi) ,\;\;\;\;\;
 v=\sqrt{\frac{\mu^2}{\lambda}}
\nonumber
\end{equation}
\begin{equation}
D_\mu \Phi  \equiv  ( \partial_\mu + \frac{1}{2} i g
\vec{W}_\mu\cdot\vec{\tau} +
\frac{1}{2} i g' B_\mu) \Phi .
\end{equation}
$\wtd, \btd$ and $\cl_{\rm GF}$ are defined in eqs.(1), (8) and (9) and
$\cl_{\rm FP}$ is the usual Faddeev--Popov term of the	standard model.

The next step is to rescale the fields and parameters in $\cl_{\rm
SM}$, according to:
\begin{eqnarray}
B_{\mu}^b & =& Z_B^{1/2} B_\mu \nonumber\\[2mm]
\vec{W}_\mu^b & = & Z_W^{1/2} \vec{W}_\mu \nonumber\\[2mm]
\vec{\pi}^b & =& Z_\pi^{1/2} \vec{\pi} \nonumber\\[2mm]
g^b & = & Z_W^{-1/2} ( g - \delta g ) \nonumber\\[2mm]
g'^b & = & Z_B^{-1/2} ( g' - \delta g' ) \nonumber\\[2mm]
v^b & = & Z_\pi^{1/2} ( v - \delta v )\nonumber\\[2mm]
\xi_B^b & = & \xi_B ( 1 + \delta \xi_B )\nonumber\\[2mm]
\xi_W^b & = & \xi_W ( 1 + \delta \xi_W )
\end{eqnarray}
We present here just the relevant parameters for the computation of
the two and three point functions with external $\gamma, W$ and $Z$
particles that are what concerns us in this work. The effects from
the renormalization of the $\lambda$ coupling and the Higgs mass on
these two and three point functions are of higher order
in perturbation theory and can be neglected from now on. This will
not be the case for the four point functions where there are
contributions from the Higgs particle already at  tree level
\cite{HR1}.

{}From the computational point of view, the large $\mh$ limit means
that one neglects the contributions to the one-light-particle
irreducible (1LPI) Green's functions that depend on (p$/\mh)^2$ and/or
(m$/\mh)^2$ and vanish when the formal
$\mh \rightarrow \infty$ limit is taken. The 1LPI functions are, by
definition, the Green's functions with only light particles in the
external legs and where the graphs contributing to them cannot be
disconnected by cutting a single light particle line.

When computing the renormalized Green's functions to one loop in the
standard model and in the large $\mh$ limit,care must be taken since
clearly the operations of making loop integrals and taking the large
$\mh$ limit do not commute. Thus, one must first regulate the loop
integrals by dimensional regularization, then
perform the renormalization operation with some fixed prescription
and at the end take the large $\mh$ limit, with $\mh$ being the
renormalized Higgs mass. The large $\mh$ values must be bounded in
practice to the range m$^2$, p$^2 \ll \mh^2 \lsim 1$ TeV$^2$ where p$^2$
represent the available external momenta and m$^2$ the light
(renormalized) particle masses (m = M$_{\rm W}$ or M$_{\rm Z}$).

Finally, the renormalized 1LPI functions with external $\gamma$, W
and Z particles are formally defined as
\begin{equation}
\Gamma_{\rm R}^{\rm SM}(\mu) = \Gamma_0^{\rm SM} + \Gamma_{\rm C}^{\rm SM}
+ \Gamma_{\rm L}^{\rm SM}
\end{equation}
where $ \Gamma_0^{\rm SM}$ is the contribution from $\cl_{\rm
SM}$ at the tree level, $\Gamma_{\rm C}^{\rm SM}$ is the contribution
from the counterterms $\delta Z_i = Z_i-1$ and $
\Gamma_{\rm L}^{\rm SM}$ is the contribution from all the one loop graphs
of the bosonic sector of the standard model  with the heavy Higgs mass
limit to be taken as explained above.

We now focus our attention on the matching condition for relating the
two theories, the fundamental underlying theory and the effective
one. We will impose here the strongest form of matching
by requiring that all renormalized 1LPI functions
with external light particles are the same in the two theories at
scales $\mu \leq \mh$.
This matching condition is
equivalent to the equality of the light particle effective action in
the two descriptions. In contrast, the weaker form of matching is
established when the equality of the two theories is done at the
physical amplitudes level.  In this work we perform the matching of
the two theories at the one loop level and we apply it to the complete
set of 1LPI functions with external $\gamma, W$ and $Z$ particles.
This matching procedure  is summarized in the following simple
condition:
\begin{equation}
\Gamma_{\rm R}^{\rm	SM}(\mu ) =\Gamma_{\rm R}^{\rm EChL}(\mu )
\; , \;\;\;\;\;\;\;\;\;\;\;\;\mu\leq \mh
\end{equation}
where the large Higgs mass limit in the left-hand
side must be understood throughout.
This equation represents symbolically a whole system of tensorial
coupled equations (as many as 1LPI functions) with several unknowns,
namely the complete set of parameters $a_i(\mu)$
and counterterms that we are interested in determining.

In the following we will present the results for the two and three
1LPI functions with external gauge particles. The matching conditions
are summarized by the following system of six tensorial equations:
\begin{eqnarray}
\Pi^{ab}_{0 \mu\nu} + \Pi^{ab}_{{\rm C} \mu\nu} +
\Pi^{ab}_{{\rm L} \mu\nu} & = &
\widehat{\Pi}^{ab}_{0 \mu\nu} + \widehat{\Pi}^{ab}_{{\rm C} \mu\nu} +
\widehat{\Pi}^{ab}_{{\rm L} \mu\nu} \label{PM}\\[2mm]
V^{abc}_{0 \lambda\mu\nu} + V^{abc}_{{\rm C} \lambda\mu\nu} +
V^{abc}_{{\rm L}\lambda\mu\nu} & = &
\widehat{V}^{abc}_{0 \lambda\mu\nu} + \widehat{V}^{abc}_{{\rm C}
 \lambda\mu\nu} +\widehat{V}^{abc}_{{\rm L} \lambda\mu\nu} \label{VM}
\end{eqnarray}
where $ab = WW, ZZ, \gamma\gamma , \gamma Z $ and $abc = \gamma WW,
ZWW$ and the large Higgs mass limit in the left-hand side must be
understood throughout.	Here we have used the hat to denote the
quantities in the effective theory, thus, the left hand side of the
above equations refers to the  standard model predictions and
the right hand side to the EChL predictions.

The resulting expressions for the tree level plus counterterms
contributions have been collected in Appendix B.
The one loop contributions to the 1LPI functions are represented in
figures 1, 2 and 3.  As mentioned above, a 1LPI function does not
include diagrams that can be disconnected by cutting a light particle
line, namely, a non-Higgs particle line.  We have included all the
one-loop 1LPI diagrams and analized them one by one in the large
$\mh$ limit and in a generic $R_\xi$--gauge.  In order to study
the large $\mh$ limit of the various one-loop Feynman
integrals, we have used the techniques developed in \cite{CAR,O}.
Particularly useful is the application of the m-theorem \cite{CAR}
and the Lebesgue dominated convergence theorem, that allows us to
discard convergent integrals that vanish in the large $\mh$ limit.
Thanks to these techniques, we have been able to perform a systematic
computation in a generic $R_\xi$--gauge that otherwise would have
been much more tedious.

We start by reporting the results on the two point functions. Fig.(1)
shows a comparative list of the one-loop 1LPI diagrams both in the
standard model and in the effective EChL theory for $WW$ and $ZZ$ two
point functions.  Among the whole set of diagrams, some of those
coming from the purely light sector  are exactly the same in the SM
and the effective EChL. This is the case of diagrams in the first
line of figs.(1.a) and (1.b), as for instance the two point diagrams
with only gauge particles flowing in the loop, (i) = ($\hat{\rm i}$),
(r) = ($\hat{\rm r}$). This subset of diagrams are represented symbolically
by ellipsis in fig.(1) and their contribution can be simply dropped
out from both sides of the matching condition (\ref{PM}).  There are
a second class of diagrams that also come from the purely light
sector, but they are not the same in the two theories, as the case of
diagrams with GB particles flowing in the loop.  This is because some
vertices in $\nll$, like those for two gauge bosons--two GB,
are different from the standard model ones, as a consequence of the
non-linear realization of the gauge symmetry. However, these long
distance contributions drop out from the matching condition because
the following identities hold in the large $\mh$ limit: (j) + (k) =
0, (s) + (t) = ($\hat{\rm s}$) and  (u) + (v) = 0.  The rest of the
tadpole diagrams with a Higgs propagator attached to the gauge line
not appearing explicitely in fig.(1) vanish in the large $\mh$ limit.
Finally, there is a third class of diagrams in the SM that contain
the Higgs particle in the loops and whose contributions to the
matching equations in the large $\mh$ limit are the relevant ones.
These are diagrams (l), (m), (n) and (o) of Fig.(1.a) and (w),
(x), (y) and (z) of Fig.(1.b).	The role being played by them in the
SM is replaced by the $a_i$ terms in the effective theory. This is an
important point, and is equivalent to saying that the difference
between the Green's functions of the two theories is an analytic
function, which for momenta much lower than the scale $\mh$ can be
approximated by a polynomial whose coefficients are given by the
$a_i$'s.

The matching conditions for the $\gamma \gamma$ and $\gamma$Z two
point functions are easily fixed since in these two cases there are
neither one-loop contributions from the Higgs particle nor any
difference in the diagrams of the two theories from the light sector.
Thus, the possible difference in the Green's functions of the two
theories are summarized in the different tree level and counterterms
contributions that can be found in Appendix B.

The final results for the loop contributions $\Pi^{ab}_{{\rm L} \mu\nu}$,
$ab = WW, ZZ, \gamma\gamma, \gamma Z$, are summarized in Appendix B.

Finally, we report the results on the three point 1LPI functions. The
diagrams contributing to the SM functions $V^{\gamma
WW}_{{\rm L} \lambda\mu\nu}$ and $V^{ZWW}_{{\rm L} \lambda\mu\nu}$ are shown in
figs.(2) and (3) respectively. The diagrams from the purely light
sector that are equal in the two theories are not shown explicitely.
As in the case of the two point functions, these contributions will
be dropped out from both sides of the matching equation (\ref{VM}).
For the rest of the diagrams the situation is the following. In the case
of $V^{\gamma WW}_{{\rm L} \lambda\mu\nu}$ in fig.(2), diagrams (b),
(c), (d), (g), (h) and (i) vanish in the large $\mh$ limit.  Besides,
it turns out that (j) + (k) =0. We have checked that the
corresponding diagram in the effective theory also vanish ($\hat{\rm
j}$)=0. In summary, we are left with three relevant diagrams, namely
(a), (e) and (f) that we have evaluated in the large $\mh$ limit.

Regarding $V^{ZWW}_{{\rm L} \lambda\mu\nu}$ in fig.(3) the situation is quite
similar.  The diagrams vanishing in the large $\mh$ limit are the
following: (b), (c), (d), (g), (h), (i), (j), (m), (n) and (o). There
is again a cancellation among diagrams, (p) + (q) =0 and corresponds
to the vanishing in the effective theory of the corresponding diagram
($\hat{\rm p}$) =0.  The relevant diagrams that we have computed in
the large $\mh$ limit are therefore (a), (e), (f), (k) and (l).

The final results for the loop contributions
$V^{VWW}_{{\rm L} \lambda\mu\nu}$, $V=\gamma ,Z$ are summarized in
Appendix B.

Having all the pieces entering in the matching
eqs.(\ref{PM},\ref{VM}) well defined, we now
solve the system formed by these six tensorial
equations. There is just one compatible solution and is given by a
set of particular values of the $a_i$ parameters and counterterms.
Obviously, the renormalized parameters $a_i(\mu)$ are scheme
renormalization dependent.  Here we have chosen the $\overline{MS}$
scheme by fixing the counterterms $\delta a_i$ to the particular
values given below. From now on, we will use the short notation
$a_i^\msb(\mu)$ for the renormalized parameters in the
$\overline{MS}$ scheme.  After some algebra we find the following
final result:
\begin{equation}
\begin{array}{rlrl}
a_0^\msb(\mu)  = & {\displaystyle g'^2  \frac{1}{16 \pi^2}
\frac{3}{8}\left( \frac{5}{6} - \log\frac{M_H^2}{\mu^2} \right)}, \;\;\;\;\;\;
\;\;\;\;& \delta a_0 = & {\displaystyle g'^2 \frac{1}{16 \pi^2} \frac{3}{8}
\Delta_\epsilon } \\[3mm]
a_1^\msb(\mu)  = & {\displaystyle \frac{1}{16 \pi^2}  \frac{1}{12}
\left( \frac{5}{6} - \log\frac{M_H^2}{\mu^2} \right)},
& \delta a_1 = & {\displaystyle \frac{1}{16 \pi^2} \frac{1}{12}
\Delta_\epsilon }\\[3mm]
a_2^\msb(\mu)  = & {\displaystyle \frac{1}{16 \pi^2} \frac{1}{24}
\left( \frac{17}{6} - \log\frac{M_H^2}{\mu^2} \right)},
& \delta a_2 = & {\displaystyle \frac{1}{16 \pi^2} \frac{1}{24}
\Delta_\epsilon }\\[3mm]
a_3^\msb(\mu)  = & {\displaystyle \frac{-1}{16 \pi^2} \frac{1}{24}
\left( \frac{17}{6} - \log\frac{M_H^2}{\mu^2} \right)},
& \delta a_3 = & {\displaystyle \frac{-1}{16 \pi^2} \frac{1}{24}
\Delta_\epsilon }\\[3mm]
a_8^\msb(\mu)  = & 0, & \delta a_8 = & 0\\[3mm] a_9^\msb(\mu)
= & 0, & \delta a_9 = & 0 \\[3mm] a_{11}^\msb(\mu)  = &{\displaystyle
\frac{-1}{16 \pi^2}\frac{1}{24}}, & \delta a_{11} = & 0 \\[3mm]
a_{12}^\msb(\mu)  = & 0, & \delta a_{12} = & 0 \\[3mm]
a_{13}^\msb(\mu)  = & 0, & \delta a_{13} = & 0 \\[3mm]
\end{array}
\label{aMH}
\end{equation}
where
\begin{equation}
\Delta_\epsilon  \equiv  \frac{2}{\epsilon} - \gamma_E + \log 4 \pi
\end{equation}

The solution (\ref{aMH}) of the matching conditions reproduces some
partial results obtained before by other authors with different
methods.  Firstly, we recover the values of the counterterms that were
first computed by Longuitano in the non-linear sigma model \cite{L}.
We obtain as well the proper logarithmic running with the scale of the
renormalized parameters $a_i(\mu)$. This $\mu$-scale dependence can
be summarized by the following set of renormalization group equations:
\begin{eqnarray}
a_0(\mu) & = &a_0(\mu') + \frac{g'^2}{16 \pi^2} \frac{3}{8}
\log\frac{\mu^2}{\mu'^2}\nonumber\\[2mm]
a_1(\mu) & = &a_1(\mu') + \frac{1}{16 \pi^2} \frac{1}{12}
\log\frac{\mu^2}{\mu'^2}\nonumber\\[2mm]
a_2(\mu) & = &a_2(\mu') + \frac{1}{16 \pi^2} \frac{1}{24}
\log\frac{\mu^2}{\mu'^2}\nonumber\\[2mm]
a_3(\mu) & = &a_3(\mu') - \frac{1}{16 \pi^2} \frac{1}{24}
\log\frac{\mu^2}{\mu'^2}
\label{RGE}
\end{eqnarray}
The paremeters $a_8$, $a_9$, $a_{11}$, $a_{12}$ and $a_{13}$ are
obviously $\mu$-independent.

It is important to stress at this point that the logarithmic running
with the $\mu$-scale depends only on the form of the lowest order universal
Lagrangian $\nll$ and therefore, it will be the same for any
underlying theory having the electroweak chiral Lagrangian as low
energy effective theory.  The differences amongst alternative
fundamental theories will come, at one loop level, in the finite
contributions to the effective Lagrangian parameters.

The rest of $a_i$'s cannot be obtained from this computation since
they do not contribute to the two and three pont functions for gauge
fields.  A completely analogous computation of the whole set of four
point 1LPI functions must be performed to extract the values of $
a_4, a_5, a_6, a_7$ and $ a_{10}$ as well as the corresponding
counterterms \cite{HR1}.

As far as the gauge sector is concerned, the matching equations
provide a set of consistency relations among the corresponding
counterterms of the two theories. We get the following relations:
\begin{eqnarray}
 \Delta Z_W & = & - \frac{g^2} {16\pi^2}
\frac{1}{12} \left(\Delta_\epsilon + \frac{5}{6}
- \log\frac{M_H^2}{\mu^2} \right)\nonumber\\[2mm]
 \Delta Z_B & = & - \frac{g'^2} {16\pi^2}
\frac{1}{12} \left(\Delta_\epsilon + \frac{5}{6}
- \log\frac{M_H^2}{\mu^2} \right)\nonumber\\[2mm]
\Delta\xi_W & = &\Delta Z_W\nonumber\\[2mm]
 \Delta\xi_B & = &\Delta Z_B\nonumber\\[2mm]
 \frac{\Delta g}{g} & = & 0\nonumber\\[2mm]
 \frac{\Delta g'}{g'} & = & 0\label{Zs}
\end{eqnarray}
where we have defined the differences of the conterterms as,
\begin{equation}
\Delta Q \equiv \delta Q- \widehat{\delta Q},\;\;\;\;\;Q=Z_W,Z_B,\xi_W,
     \xi_B,g,g'\nonumber
\end{equation}
The wave function renormalization constants are not the same in the
two theories, a result that accounts for the fact that the effective
theory must incorporate the additional logarithmic divergences that
arise in the  standard model only when $\mh\rightarrow \infty$.
Our results of eq.(\ref{Zs}) confirm those found by Longhitano in
\cite{L} that referred just to the terms proportional to
$\Delta_\epsilon$.  The results in eqs.(\ref{Zs}) give a set of
constraints relating the renormalization prescriptions for the
effective and the underlying theory. These equations tell us the
way one must fix the wave functions and coupling constants
renormalizations in the effective theory, once a particular
prescription for the standard model counterterms has been assumed.

Finally, to end this section, we have performed a comparison between
our results of eq.(\ref{aMH}) and the chiral parameters found by
Gasser and Leutwyler in the first paper in \cite{GL} for the case of
the linear  sigma model with spontaneously broken $SU(2)_{\rm L}\times
SU(2)_{\rm R}$ symmetry. Since in their case there is no custodial symmetry
breaking our set of operators defined in eq.(\ref{Li}) is larger and
contains theirs. A simple exercise shows that the relation among the
two sets of bare parameters is the following:
\begin{equation}
L_{10}=a_1 ,\;\;\;\; L_9=a_3-a_2 ,\;\;\;\;L_1=a_5 ,\;\;\;\;L_2=a_4 .
\end{equation}

\noindent
After performing the renormalization operation in both the
$\msb$ scheme and the Gasser and Leutwyler ($GL$) scheme
which differ in a finite constant, we find:
\begin{eqnarray}
L_{10}^{\rm GL}(\mu) & = & L_{10}^\msb(\mu) - \frac{1}{16\pi^2}
\frac{1}{12} = a_1^\msb(\mu) - \frac{1}{16\pi^2} \frac{1}{12}=
 \nonumber\\[2mm]
  &=&
\frac{-1}{16\pi^2}
\left(\frac{1}{72}+\frac{1}{12}\log \frac{M_H^2}{\mu^2}\right)
\nonumber\\[2mm]
L_9^{\rm GL}(\mu) & = & L_9^\msb(\mu) + \frac{1}{16\pi^2} \frac{1}{12}=
a_3^\msb(\mu) - a_2^\msb(\mu) + \frac{1}{16\pi^2} \frac{1}{12}=
\nonumber\\[2mm]
&=&
\frac{-1}{16\pi^2}
\left(\frac{11}{72}-\frac{1}{12}\log \frac{M_H^2}{\mu^2}\right)
\end{eqnarray}
These values agree with the results found by Gasser and Leutwyler in
\cite{GL}. We find this a quite remarkable result since in their case
there are no gauge particles in the loops because the gauge fields
were considered as external sources.
We believe that this result can be traced back to the fact that
the contributions to the effective operators
of dimension four that come from mixed gauge--scalar loops are
subleading, in the large $\mh$ limit, as compared to the pure
scalar loops contributions.
On the other hand,  since the custodial breaking operators
are generated precisely by these mixed loops, one can conclude
that the dimension four custodial breaking operators do not
get contributions from the Higgs particle at one loop.
However, this is not the case for dimension two operators.
The  custodial breaking operator corresponding to $a_0$ comes
from mixed gauge-scalar loops in diagrams (m) in fig.(1.a) and (x)
in fig.(1.b) which give a non-vanishing contribution to $a_0$
in the large $\mh$ limit.

\section{Some physical consequences}

We would like to add in this section some remarks and comments on
the results for the EChL parameters presented in eq.(\ref{aMH}).
These finite values represent the non-decopling effects of a heavy
Higgs particle in the SM to one-loop order. They contain valuable
information since they serve as reference values to be compared with
the corresponding predictions from  other possible alternatives for
the symmetry breaking sector. Thus, for instance, in Technicolor
Models, the values for the $a_i$ parameters are known to be quite
different \cite{AW}. The optimal strategy will be therefore to find a
set of appropriate observables that, once expressed in terms of the
$a_i$'s, can provide a systematic check of the compatibility of the
assumed underlying theory to one-loop level with data. Some of these
observables like the $T,U$ and $S$ parameters of Peskin and Takeuchi
\cite{PT} or the related parameters $\epsilon_1$,$\epsilon_2$ and
$\epsilon_3$ of Altarelli and Barbieri
\cite{AlB} have already been studied by many authors in conexion with the
LEP data \cite{HDG}. The contributions from the $a_i$'s to the
$\epsilon$ parameters are given by \cite{F}:
\begin{eqnarray}
\epsilon_1 & = & 2 a_0^\msb(M_Z)\nonumber\\[2mm]
\epsilon_2 & = & - g^2 \left( a_8^\msb(M_Z) +
a_{13}^\msb(M_Z) \right) \nonumber\\[2mm]
\epsilon_3 & = & - g^2 \left( a_1^\msb(M_Z) +
a_{13}^\msb(M_Z) \right)
\end{eqnarray}
where we have chosen $M_Z$ as the reference low energy scale.

{}From the present work, therefore, we are able to compute the
contribution from the Higgs particle in the SM to one loop and in the
large $\mh$ limit to the values of the $\epsilon$ parameters. From
eqs.(\ref{aMH}) and (\ref{RGE}) we get:
\begin{eqnarray}
\epsilon_1 & = & \alpha T = \Delta \rho =
\frac{g'^2} {16\pi^2} \left( \frac{15}{24} -
\frac{3}{4} \log\frac{M_H^2}{M_Z^2}\right)+...\nonumber\\[2mm]
\epsilon_2 & = & \frac{-\alpha}{4s_W^2} U = 0+...\nonumber\\[2mm]
\epsilon_3 & = & \frac{\alpha}{4s_W^2} S =
\frac{g^2}{16\pi^2} \left( \frac{-5}{72} +
\frac{1}{12} \log\frac{M_H^2}{M_Z^2}\right)+...
\label{eps}
\end{eqnarray}
The leading logarithmic terms agree with previous computations in the
literature \cite{CH}. The next to leading terms in the large $\mh$ limit
 are independent on $\mh$ and have been computed here for the first
time. The dots in eq.(\ref{eps}) refer to the rest of the loop
contributions in the SM other than the Higgs contributions.

It is interesting to
notice that there are certain particular combinations of observables
that are independent on the choice of the reference energy scale.
Similarly one can say that these combinations are renormalization group
invariants in the effective theory and may have some relevance in the
search of physical effects beyond the standard model
\footnote{This issue was
discussed firstly in the second reference of \cite{HDG} and in
\cite{EH} where the renormalization group invariants were called
$O_1$, $O_2$ and $O_3$.}. One of these
combinations in terms of the $\epsilon$'s is
$\displaystyle{\left(\epsilon_1+9\frac{g'^2}{g^2}\epsilon_3\right)}$.
By substituting the values of
 eq.(\ref{eps}) we find that there are no Higgs contributions to
this   particular  combination, namely, both contributions the
logarithmic ones and the constant terms cancel,
$\displaystyle{\left(\epsilon_1+9\frac{g'^2}{g^2}\epsilon_3\right)=0+..}$.
We believe it is an interesting result since by means of these combinations
one can better isolate the
effects from possible alternatives to the symmetry breaking sector of
the standard model or, more  generally, from possible new physics
beyond the standard model\footnote{We refer the reader to
ref.\cite{EM} for further discussions on the relevance of finding
renormalization group invariant quatities in Chiral perturbation
theory}.

The next generation of interesting observables are the parameters
defining possible deviations of the trilinear gauge boson vertex.
These are the usual anomalous couplings $g_1^{\gamma}$ and $g_1^Z$
and anomalous magnetic moments of the $W$, $\kappa_{\gamma}$,
$\kappa_Z$, $\lambda_{\gamma}$ and $\lambda_Z$ which have been object
of numerous studies in the past\footnote{For some recent discussions on
this subject, in the context of effective lagrangians and, in connection
with the LEP II experiment see for instance \cite{EH,DeR}}.
 The contributions
from the non-vanishing $a_i$'s to these parameters are \cite{F}:
\begin{eqnarray}
g_1^{\gamma} - 1 & = & 0\nonumber\\[2mm] g_1^Z - 1 & = & \frac{-
g^2}{c_W^2} a_3^\msb(\mu)\nonumber\\[2mm]
\kappa_{\gamma} - 1 & = & g^2 \left( a_2 - a_3 - a_1 \right)\nonumber\\[2mm]
\kappa_Z - 1 & = & - g^2 a_3^\msb(\mu) +
g'^2 \left( a_1^\msb(\mu) - a_2^\msb(\mu) \right)\nonumber\\[2mm]
\lambda_{\gamma} = \lambda_Z & = & 0
\end{eqnarray}
where $\mu$ is the appropiate scale to be fixed according to the
relevant energy scale of the experiment where these parameters will
be meassured.

One particularly interesting result is the combination of parameters
entering in the definition of $\kappa_{\gamma}$ being $\mu$-scale
independent. Our prediction for this renormalization group invariant
is:
\begin{equation}
\kappa_{\gamma}-1=\frac{1}{16\pi^2}\frac{1}{6}+... \label{kp}
\end{equation}

\noindent
where, as before, the dots refer to the rest of the loop contributions
other than the Higgs contributions. This value is in agreement with the
result found in
\cite{BGL} a long
time ago once the large $\mh$ limit is taken in their expressions. We
find it to be a good check of our computation being performed in
a completely independent and quite different way. In particular the
computation of ref.\cite{BGL} was performed in the unitary gauge.

Futhermore, we believe this observable is also of interest because it
appears in the amplitude for the scattering process $\gamma\gamma
\rightarrow W^+_{\rm L} W^-_{\rm L}$ and therefore if the future planned
dedicated
$\gamma \gamma$ colliders are carried out it could be meassured with
a good precission \cite{PP}. This process was computed in the SM to
one-loop in \cite{BJ} and with the EChL effective approach in
\cite{HR2}. After taking the large $\mh$ limit of the amplitudes in
\cite{BJ}, we obtain exactly the same value for $\kappa_{\gamma}$ of
eq.(\ref{kp}). This second check is also remarkable since the authors
of ref.\cite{BJ} used the Feynman-t'Hooft gauge, the equivalence
theorem, and worked with the on-shell renormalization prescription.

\section{Conclusions}

In this paper, we have considered the electroweak chiral
Lagrangian that parametrizes electroweak interactions in the case of
a heavy Higgs boson. This Lagrangian has been obtained as the low
energy effective theory of the standard model when the Higgs particle
is integrated out to one--loop order. In particular, we have
analyzed the subset of effective operators that contribute to the
two--  and three--point Green's functions for gauge bosons. The
leading contributions of a heavy Higgs to these operators, including
logarithms of the Higgs mass plus finite (non-logarithmic) terms, have
been explicitely calculated.

The electroweak chiral Lagrangian provides a general framework to
analyze the effects of alternative dynamics of the Higgs sector in
low energy observables. Therefore, it is interesting to determine the
parameters in the case of a heavy Higgs, since they serve as
reference values to be compared with those coming from other models
of symmetry breaking. The chiral parameters are directly related to
different observables in scattering processes and in precision
electroweak measurements,
and therefore can be used to constrain the underlying dynamics from
experimental data.

\section*{Aknowledgments}
We would like to thank A.Dobado and D.Espriu for reading the manuscript
and for useful discussions. A special thank goes to C.P.Mart\'\i n for
many stimulating discussions, his interesting comments on the manuscript
and for his valuable help and patience
in explaining us the large-m techniques developed in
\cite{CAR,O}. We are also indebted to C.Quimbay who participated in
the very early stages of this work. M.J.H. aknowledges partial
finantial support
from the Ministerio de Educacion y Ciencia (Spain) (CICYT AEN90-272).
E.R.M. aknowledges support from Dpto. de Matem\'aticas at UAM
and thanks the SLAC theory group for their hospitality.

\newpage
\section*{Appendix A}
The complete expression for $\fpnl$ containing terms up to
$O(c c \pi \pi )$ is:
\begin{eqnarray}
\fpnl & = & c_0^{\dagger} \left[-\nabla^2 -g'\left( \frac{g' v \xi_B}{2}
\right)
\left[\frac{v}{2}-\frac{1}{6 v}(\pi_1^2 + \pi_2^2) + ...\right] \right]
c_0 \nonumber\\
& & +\sum_{i\ne j \ne k = 1}^3 c_i^{\dagger} \left[-\nabla^2 -g\left(
\frac{g
 v \xi_W}{2} \right)
\left[\frac{v}{2}-\frac{1}{6 v}(\pi_j^2 + \pi_k^2) + ...\right] \right]
c_i \nonumber\\
& & + ( g'\xi_B c^{\dagger}_0 c_1 + g\xi_W c^{\dagger}_1 c_0)
\left(\frac{\sqrt{g g'} v}
{2} \right) \left[ \frac{- \pi_2}{2} + \frac{1}{6 v} \pi_3 \pi_1 + ...\right]
\nonumber\\
& & + ( g'\xi_B c^{\dagger}_0 c_2 + g \xi_W c^{\dagger}_2 c_0)
\left(\frac{\sqrt{g g'} v}
{2} \right) \left[ \frac{ \pi_1}{2} + \frac{1}{6 v} \pi_3 \pi_2 + ...\right]
\nonumber\\
& & + ( g' \xi_B c^{\dagger}_0 c_3 + g \xi_W c^{\dagger}_3 c_0)
\left(\frac{\sqrt{g g'} v}
{2} \right) \left[ \frac{v}{2} - \frac{1}{6 v}(\pi_1^2 + \pi_2^2) + ...\right]
\nonumber\\
& & + (c^{\dagger}_1 c_2 - c^{\dagger}_2 c_1) \left[ - g \partial^\mu
W^3_\mu + g \left(\frac{g v \xi_W}{2}\right) \frac{\pi_3}{2} \right]
\nonumber\\
& & + (c^{\dagger}_1 c_3 - c^{\dagger}_3 c_1) \left[  g \partial^\mu
W^2_\mu - g \left(\frac{g v \xi_W}{2}\right) \frac{\pi_2}{2} \right]
\nonumber\\
& & + (c^{\dagger}_2 c_3 - c^{\dagger}_3 c_2) \left[ - g \partial^\mu
W^1_\mu + g \left(\frac{g v \xi_W}{2}\right) \frac{\pi_1}{2} \right]
\nonumber\\
& & + (c^{\dagger}_1 c_2 + c^{\dagger}_2 c_1) g \left(\frac{g v
\xi_W}{2}\right)
\left(\frac{-1}{6 v}\right) \pi_1 \pi_2 +...\nonumber\\
& & + (c^{\dagger}_1 c_3 + c^{\dagger}_3 c_1) g \left(- \frac{g v
\xi_W}{2}\right)
\left(\frac{+1}{6 v}\right) \pi_1 \pi_3 +...\nonumber\\
& & + (c^{\dagger}_2 c_3 + c^{\dagger}_3 c_2) g \left(\frac{g v
\xi_W}{2}\right)
\left(\frac{-1}{6 v}\right) \pi_2 \pi_3 +...\nonumber
\end{eqnarray}
\vspace{1cm}

\section*{Appendix B}

In this appendix, we give the different contributions to the
one loop Green's functions appearing in the matching equations
(\ref{PM},\ref{VM}). In these formulas, we have not written
explicitely the counterterm in the bare $a_i$ coefficients for
brevity; therefore the replacement $a_i \rightarrow a_i^{\msb}(\mu)
+ \delta a_i^{\msb}$ has to be understood.
The rest of fields and parameters are renormalized quantities.

\subsection*{Two point functions}
The differences in the tree level plus counterterm contributions to the
two point functions from the SM and the EChL are given by:

\begin{eqnarray}
-i \Pi^{WW}_{(0+{\rm C}) \mu\nu}+i \widehat{\Pi}^{WW}_{(0+{\rm C})
\mu\nu}&=&
i g_{\mu\nu} \frac{g^2 v^2}{4}
\left[
\Delta Z_\pi -
2 \frac{\Delta g}{g} - 2 \frac{\Delta v}{v} \right]
\nonumber \\[2mm]
& & -i \left( g_{\mu\nu} q^2 - q_\mu q_\nu \right) \Delta Z_W
+ i q_\mu q_\nu \left[\frac{1}{\xi_W} (\Delta\xi_W-\Delta Z_W )
+ g^2 a_{11}\right] \nonumber \\[4mm]
-i \Pi^{ZZ}_{(0+{\rm C}) \mu\nu} +i \widehat{\Pi}^{ZZ}_{(0+{\rm C})
\mu\nu}&=&
i g_{\mu\nu}\frac{(g^2+g'^2) v^2}{4}
\left[\Delta Z_\pi-
2c_{\rm w}^2\frac{\Delta g}{g}-2s_{\rm w}^2\frac{\Delta g'}{g'}
-2\frac{\Delta v}{v}+2a_0\right] \nonumber \\[2mm]
& & -i \left( g_{\mu\nu} q^2 - q_\mu q_\nu \right)
\left[ c_{\rm w}^2 \Delta Z_W
+ s_{\rm w}^2 \Delta Z_B \frac{}{}\right. \nonumber\\[2mm]
& & \left. - c_{\rm w}^2 g^2 a_8 - 2 s_{\rm w}^2 g^2 a_1
- (g^2+g'^2) a_{13} \frac{}{} \right] \nonumber\\[2mm]
& &  +i q_\mu q_\nu \left[ \frac{c_{\rm w}^2}{\xi_W} \left(
\Delta \xi_W - \Delta Z_W \right) + \frac{s_{\rm w}^2}{\xi_B}
\left( \Delta \xi_B - \Delta Z_B \right) \right. \nonumber\\[2mm]
& & \left. + \left(g^2 + g'^2 \right) \left( a_{11} - 2 a_{12} +
a_{13} \right) \frac{}{}\right] \nonumber\\[4mm]
-i \Pi^{\gamma\gamma}_{(0+{\rm C}) \mu\nu} +i
\widehat{\Pi}^{\gamma\gamma}_{(0+{\rm C}) \mu\nu}&=&
-i \left( g_{\mu\nu} q^2 - q_\mu q_\nu \right) \left[
s_{\rm w}^2 \Delta Z_W +
c_{\rm w}^2 \Delta Z_B -
s_{\rm w}^2 g^2 (a_8 - 2 a_1)\frac{}{}\right]\nonumber \\[2mm]
& &+i q_\mu q_\nu \left[ \frac{s_{\rm w}^2}{\xi_W} \left(
\Delta \xi_W - \Delta Z_W \right) + \frac{c_{\rm w}^2}{\xi_B}
\left( \Delta \xi_B - \Delta Z_B \right) \right] \nonumber\\[4mm]
-i \Pi^{\gamma Z}_{(0+{\rm C}) \mu\nu} +i \widehat{\Pi}^{\gamma
Z}_{(0+{\rm C}) \mu\nu}&=&i g_{\mu\nu}	\frac{g g' v^2}{4}
 \left[ \frac{\Delta g'}{g'} - \frac{\Delta g}{g}
\right] \nonumber\\[2mm]
& & -i \left( g_{\mu\nu} q^2 - q_\mu q_\nu \right) \left[
s_{\rm w} c_{\rm w} \Delta Z_W -
s_{\rm w} c_{\rm w} \Delta Z_B \frac{}{}\right.\nonumber\\[2mm]
& & \left. - s_{\rm w} c_{\rm w} g^2 a_8 +
 (c_{\rm w}^2 - s_{\rm w}^2) g g' a_1 \frac{}{}\right] \nonumber\\[2mm]
& & +i q_\mu q_\nu s_{\rm w} c_{\rm w} \left[ \frac{1}{\xi_W} \left(
\Delta \xi_W - \Delta Z_W \right) - \frac{1}{\xi_B}
\left( \Delta \xi_B - \Delta Z_B \right) \right] \nonumber
\end{eqnarray}

\noindent
In the expressions above, the $\Delta$ quantities represent the
standard model counterterms minus the corresponding counterterms in the
EChL, that is
\vspace{4mm}

$
\hspace{1cm}\Delta Q \equiv \delta Q - \widehat{\delta Q}, \hspace{8mm}
{\rm with} \hspace{8mm} Q = Z_{\pi},\; Z_B,\; Z_W,\; \xi_B,\; \xi_W,\;
g,
\; g'\;\;
{\rm and} \;\; v.$
\vspace{4mm}

\noindent
On the other hand, the one loop contributions wich do not cancel among both
sides of the matching condition (\ref{PM}) are the following:
\begin{eqnarray}
-i \Pi^{WW}_{{\rm L} \mu\nu} +i \widehat{\Pi}^{WW}_{{\rm L} \mu\nu}
& = & i g_{\mu \nu} \frac{g^2}{16 \pi^2} \left[  \frac{3}{4} M_H^2
\left( \Delta_\epsilon - \log\frac{M_H^2}{\mu^2} + \frac{7}{6} \right)
\right. \nonumber\\[2mm]
& & \left. - \frac{g^2 v^2}{4} \frac{3}{4}
\left( \Delta_\epsilon - \log\frac{M_H^2}{\mu^2} + \frac{5}{6} \right)
\right] \nonumber\\[2mm]
& & - i \left( g_{\mu\nu} q^2 - q_\mu q_\nu \right) \frac{g^2}{16 \pi^2}
\frac{1}{12}
\left( \Delta_\epsilon - \log\frac{M_H^2}{\mu^2} + \frac{5}{6} \right)
\nonumber\\[2mm]
& & + i q_\mu q_\nu \; \frac{g^2}{16 \pi^2} \frac{1}{24} \nonumber  \\[4mm]
-i \Pi^{ZZ}_{{\rm L} \mu\nu} +i \widehat{\Pi}^{ZZ}_{{\rm L} \mu\nu}
& = & i g_{\mu \nu} \frac{(g^2 + g'^2)}{16 \pi^2} \left[
\frac{3}{4}  M_H^2 \left( \Delta_\epsilon
- \log\frac{M_H^2}{\mu^2} + \frac{7}{6} \right) \right. \nonumber\\[2mm]
& & \left. - \frac{(g^2 + g'^2) v^2}{4} \frac{3}{4}
\left( \Delta_\epsilon - \log\frac{M_H^2}{\mu^2} + \frac{5}{6} \right)
\right] \nonumber\\[2mm]
& & - i \left( g_{\mu\nu} q^2 - q_\mu q_\nu \right)
\frac{(g^2 + g'^2)}{16 \pi^2}\frac{1}{12}
\left( \Delta_\epsilon - \log\frac{M_H^2}{\mu^2} + \frac{5}{6} \right)
\nonumber\\[2mm]
& & + i q_\mu q_\nu \frac{(g^2 + g'^2)}{16 \pi^2}  \frac{1}{24}
\nonumber\\[4mm]
-i \Pi^{\gamma\gamma}_{{\rm L} \mu\nu} +i \widehat{\Pi}^{\gamma\gamma}_
{{\rm L} \mu\nu} & = & 0 \nonumber\\[4mm]
-i \Pi^{\gamma Z}_{{\rm L} \mu\nu} +i \widehat{\Pi}^{\gamma Z}_
{{\rm L} \mu\nu}
& = & 0 \nonumber
\end{eqnarray}

\subsection*{Three point functions}
The differences in the tree level plus counterterm contributions to the
$\gamma$WW and ZWW 1PI Green's functions from the SM and the EChL are
given by:
\begin{eqnarray}
-i V^{\gamma WW}_{(0+{\rm C}) \lambda\mu\nu}
+i \widehat{V}^{\gamma WW}_{(0+{\rm C}) \lambda\mu\nu}
& = & -i g s_{\rm w} T^\gamma_{\lambda\mu\nu}
\left[
\Delta Z_W - \frac{\Delta g}{g}
\right] \nonumber \\[2mm]
& &  - i g^3 s_{\rm w} \left( p_{1 \mu} g_{\lambda \nu} - p_{1 \nu}
g_{\lambda \mu}
\right) \left[ a_1 - a_2 + a_3 - a_8 + a_9 \frac{}{}\right] \nonumber\\[2mm]
& & -i g^3 s_{\rm w}  \left( p_{2 \mu} g_{\lambda \nu} - p_{3 \nu}
g_{\lambda \mu}
 \right) a_{11} \nonumber\\[4mm]
-i V^{ZWW}_{(0+{\rm C}) \lambda\mu\nu}
+i \widehat{V}^{ZWW}_{(0+{\rm C}) \lambda\mu\nu} & = &
-i g c_{\rm w} T^Z_{\lambda\mu\nu}
\left[
\Delta Z_W - \frac{\Delta g}{g}
+ \frac{g^2}{c_{\rm w}^2} a_3
\right]\nonumber \\[2mm]
& &  - i g c_{\rm w} \left( p_{1 \mu} g_{\lambda \nu} - p_{1 \nu}
g_{\lambda \mu}
\right) \left[ g'^2 ( a_2 - a_1 - a_3 - a_{13} ) \frac{}{} \right.
\nonumber\\[2mm]
& & \left.+ g^2 ( a_9 - a_8 - a_{13} ) \frac{}{}\right] \nonumber\\[2mm]
& &  - i g c_{\rm w} \left( p_{2 \mu} g_{\lambda \nu} - p_{3 \nu}
g_{\lambda\mu}
\right) \left[ -g'^2 a_{11} + \frac{g^2}{c_{\rm w}^2} a_{12} \right]\nonumber
\end{eqnarray}
where $\Delta Z_W$ and $\Delta g$ are defined as in the two point
functions case. The convention for momenta and indexes in the three
point functions are defined by the tensor associated to the tree level
vertex \vspace{4mm}

$
T^{V}_{\lambda\mu\nu} \equiv \left( V_\lambda(p_1),
W^{-}_\mu(p_2), W^+_\nu(p_3) \right) = \left[ (p_1-p_3)_\mu
g_{\lambda\nu} + (p_3-p_2)_\lambda g_{\mu\nu} + (p_2-p_1)_\nu
g_{\lambda\mu} \right]$ \vspace{2mm}

\noindent
and $V = \gamma$ or $Z$. All the momenta are taken incoming.\vspace{2mm}

\noindent
The one loop contributions which enter the matching equations
(\ref{VM}) are the following:
\begin{eqnarray}
-i V^{\gamma WW}_{{\rm L} \lambda\mu\nu}
+i \widehat{V}^{\gamma WW}_{{\rm L} \lambda\mu\nu}
& = & -i g s_{\rm w} T^\gamma_{\lambda\mu\nu} \frac{g^2}{16 \pi^2} \frac{1}{12}
\left( \Delta_\epsilon - \log\frac{M_H^2}{\mu^2} + \frac{5}{6}
\right) \nonumber\\[2mm]
& &  + i g s_{\rm w} \left( p_{1 \mu} g_{\lambda \nu} - p_{1 \nu}
g_{\lambda \mu} \right)  \frac{g^2}{16 \pi^2} \frac{-1}{6} \nonumber\\[2mm]
& & +i g s_{\rm w}  \left( p_{2 \mu} g_{\lambda \nu} - p_{3 \nu}
g_{\lambda \mu} \right)  \frac{g^2}{16 \pi^2} \frac{-1}{24} \nonumber\\[4mm]
-i V^{ZWW}_{{\rm L} \lambda\mu\nu}
+i \widehat{V}^{ZWW}_{{\rm L} \lambda\mu\nu}
& = & -i g c_{\rm w} T^{Z}_{\lambda\mu\nu} \left[
\frac{g^2}{16 \pi^2} \frac{1}{12}
\left( \Delta_\epsilon - \log\frac{M_H^2}{\mu^2} + \frac{5}{6}\right) \right.
\nonumber\\[2mm]
& & \left. + \frac{(g^2+g'^2)}{16 \pi^2 } \frac{1}{24}
\left( \Delta_\epsilon - \log\frac{M_H^2}{\mu^2} + \frac{17}{6} \right)
\right] \nonumber\\[2mm]
& &  + i g c_{\rm w} \left( p_{1 \mu} g_{\lambda \nu} - p_{1 \nu}
g_{\lambda \mu} \right)  \frac{g'^2}{16 \pi^2} \frac{1}{6}\nonumber\\[2mm]
& & +i g c_{\rm w}  \left( p_{2 \mu} g_{\lambda \nu} - p_{3 \nu}
g_{\lambda \mu} \right)  \frac{g'^2}{16 \pi^2} \frac{1}{24} \nonumber
\end{eqnarray}
These are all the necessary contributions to solve the matching
equations (\ref{PM}, \ref{VM}).
\newpage

\section*{Figure Captions}
\begin{description}

\item[Fig.1] {\bf 1.a} One-loop diagrams contributing to the $WW$ self-
energy in the standard model (left side) and in the effective
EChL theory (right side).\\
{\bf 1.b} Same as 1.a for the $ZZ$ self-energy.

\item[Fig.2] One-loop diagrams contributing to the $\gamma WW$ 1LPI
Green's function in the standard model that differ
from those in the EChL.

\item[Fig.3] Same as Fig.2 for the $ZWW$ three point function.

\end{description}


\newpage

\begin{thebibliography}{99}
\bibitem{W} S.Weinberg, Physica {\bf 96A} (1979), 327.
\bibitem{GL} J.Gasser and H.Leutwyler, Ann. Phys. (N.Y.) {\bf 158} (1984), 142;
Nucl. Phys. {\bf B250} (1985), 465.
\bibitem{D} A pedagogical introduction to the subject of Effective
Chiral Lagrangians can be found in:\\
J.Donoghue, E.Golowich and B.R.Holstein. Cambridge University Press,
1992.
\bibitem{DHD}The idea of using Effective Chiral Lagrangians and Chiral
Perturbation Theory in the context of the electroweak interactions was
proposed in:\\
A.Dobado and M.J.Herrero,
Phys. Lett {\bf B228} (1989),495;  {\bf B233} (1989),505.\\
J.Donoghue and C.Ramirez, Phys. Lett. {\bf B234} (1990), 361.
\bibitem{DHT}A.Dobado, M.J.Herrero and J.Terron, Z. Phys. {\bf C50}
(1991), 205, 465.\\
A.Dobado, M.J.Herrero and T.Truong, Phys. Lett. {\bf B235} (1990),129.\\
S.Dawson and G.Valencia, Nucl. Phys. {\bf B352} (1991), 27.\\
J.Barger, S.Dawson and G.Valencia, Fermilab-Pub-92/75-T,1992.\\
A.Dobado and M.Urdiales, Phys. Lett. {\bf B292} (1992), 128.
\bibitem{HDG} B.Holdom and J.Terning, Phys. Lett. {\bf B247} (1990), 88.\\
A.Dobado, D.Espriu and M.J.Herrero, Phys. Lett. {\bf B255} (1991), 405.\\
M.Golden and L.Randall, Nucl. Phys. {\bf B361} (1991), 3.
\bibitem{EH} D.Espriu and M.J.Herrero, Nucl. Phys. {\bf B373}
 (1992), 117.
\bibitem{G} H.Georgi, Nucl. Phys. {\bf B363} (1991), 301; Nucl. Phys. {\bf
29B,C} (Proc. Suppl.) (1992), 1; Nucl. Phys. {\bf B361} (1991), 339.
\bibitem{FMM} F.Ferruglio, A.Masiero and L.Maiani, Nucl. Phys. {\bf B387}
(1992), 523.
\bibitem{F} F.Ferruglio {\sl in} Lectures at the 2$^{nd}$ Nat. Seminar
of Th. Physics, Parma, Sept. 1992. DFPD92/TH/50.
\bibitem{Sint} For a review see also,\\
S.Sint, Diplomarbeit Universitat Hamburg (1991).
\bibitem{TEC} S.Weinberg, Phys. Rev. {\bf D19} (1979), 1277.\\
L.Susskind, Phys. Rev.{\bf D20} (1979), 2619.\\
E.Farhi and L.Susskind, Phys. Rep. {\bf 74} (1981), 279.
\bibitem{Lynn} B.W.Lynn, M.E.Peskin and R.G.Stuart, CERN-86-02 (1986).\\
M.E.Peskin and T.Takeuchi, Phys. Rev. Lett. {\bf 65} (1990), 964.
\bibitem{AW} T.Appelquist and G.-H Wu, YCTP-P7-93,  April 1993.
\bibitem{AB} T.Appelquist and C.Bernard, Phys. Rev. {\bf D22}
(1980),200;\\
T.Appelquist {\sl in} Gauge Theories and Experiments at High Energies,
Ed. K.C.Browner and D.G.Sutherland, Scottish U. Summer School, 1980.
\bibitem{L} A.C.Longuitano, Nucl. Phys. {\bf B188} (1981), 118;
Phys. Rev. {\bf D22} (1980), 1166.
\bibitem{VG} M.Veltman Acta Phys. Pol. {\bf B8} (1977) 475.\\
M.Lemoine and M.Veltman, Nucl.Phys.{\bf B164} (1980), 445.\\
O.Cheyette and M.K.Gaillard, Phys. Lett. {\bf B197} (1987), 205.\\
H.Veltman and M.Veltman, Acta Phys. Pol. {\bf B22} (1991), 669
\bibitem{LE} M.S.Chanowitz and M.K.Gaillard, Nucl.Phys.{\bf B261}
(1985), 379.\\
M.S.Chanowitz, M.Golden and H.Georgi, Phys.Rev.{\bf D36} (1987), 1490.
\bibitem{CAR} G.Giavarini, C.P.Martin and F. Ruiz Ruiz, Nucl. Phys.
{\bf B381} (1992), 222.
\bibitem{O} E.B. Manoukian, J. Math. Phys. {\bf 22} (3) (1981), 572; {\bf
22} (10) (1981), 2258.
\bibitem{EM} D.Espriu and J.Matias, Univ. Barcelona preprint,
June 1993, UB-ECM-PF 93/15.
\bibitem{PT}
M.E.Peskin and T.Takeuchi, Phys. Rev. {\bf D46} (1992), 381.
\bibitem{AlB} G.Altarelli and R.Barbieri, Phys. Lett. {\bf B253} (1991), 161;\\
G.Altarelly, R.Barbieri and S.Jadach, Nucl. Phys. {\bf B269} (1992), 3.
\bibitem{DeR} A. De Rujula et al., Nucl. Phys. {\bf B384} (1992), 3.\\
P.Hernandez and F.J.Vegas, CERN-TH-6670 (1992).\\
M.Bilenky et al., BI-TP-92/44 (1992).
\bibitem{BGL} W.A.Bardeen, R.Gastmans and B.Lautrup, Nucl. Phys. {\bf B46}
(1972), 319.
\bibitem{PP} I.F.Ginzburg et al. Nucl. Inst. Meths.205 (1983), 47;
219(1984), 5.\\
I.F.Ginzburg et al. Nucl. Phys. {\bf B228} (1983), 285.
\bibitem{BJ} E.E.Boos and G.V.Jikia, Phys. Lett. {\bf B275} (1992), 164.
\bibitem{HR1} M.J.Herrero and E.Ruiz Morales, work in preparation.
\bibitem{HR2} M.J.Herrero and E.Ruiz Morales, Phys. Lett. {\bf B296}
(1992), 397.
\bibitem{CH} See for instance:\\
 M.Consoli and W.Hollik in Z
Phys.
at LEP I, CERN Yellow Report, ed. G.Altarelli et al (CERN, Geneva, 1989)
\end{thebibliography}
\end{document}